\documentclass[letter]{aa}  
\usepackage[varg]{txfonts}
\usepackage{newtxtext,newtxmath}
\usepackage[T1]{fontenc}

\DeclareRobustCommand{\VAN}[3]{#2}
\let\VANthebibliography\thebibliography
\def\thebibliography{\DeclareRobustCommand{\VAN}[3]{##3}\VANthebibliography}

\usepackage{color}
\usepackage{booktabs}
\usepackage{enumitem}
\usepackage{hyperref}
\usepackage{verbatim}
\usepackage{amsmath,amstext,mathrsfs}
\usepackage[all]{hypcap} 
\usepackage{afterpage}    
\usepackage{marginnote}
\usepackage{makecell}
\usepackage{subfigure}
\usepackage{xfrac}

\usepackage{ifxetex,ifluatex}
\usepackage{fixltx2e} % provides \textsubscript
\usepackage{multirow}

\DeclareFontFamily{OMS}{oasy}{\skewchar\font48 }
\DeclareFontShape{OMS}{oasy}{m}{n}{%
         <-5.5> oasy5     <5.5-6.5> oasy6
      <6.5-7.5> oasy7     <7.5-8.5> oasy8
      <8.5-9.5> oasy9     <9.5->  oasy10
      }{}
\DeclareFontShape{OMS}{oasy}{b}{n}{%
       <-6> oabsy5
      <6-8> oabsy7
      <8->  oabsy10
      }{}
\DeclareSymbolFont{oasy}{OMS}{oasy}{m}{n}
\SetSymbolFont{oasy}{bold}{OMS}{oasy}{b}{n}

\DeclareMathSymbol{\smallleftarrow}     {\mathrel}{oasy}{"20}
\DeclareMathSymbol{\smallrightarrow}    {\mathrel}{oasy}{"21}
\DeclareMathSymbol{\smallleftrightarrow}{\mathrel}{oasy}{"24}

\graphicspath{{./}{fig/}}

\title{Correlation of velocity and density contributions to spectroscopic channel maps: Reality check on Kalberla et.al (2022)}

   \author{Ka Ho Yuen
          \inst{1}
          Ka Wai Ho\inst{1} \and
          Alex Lazarian\inst{1,2}
          }

   \institute{Department of Astronomy, University of Wisconsin-Madison, US\\
              \email{kyuen@astro.wisc.edu,kho33@wisc.edu,alazarian@facstaff.astro.wisc.edu}
              \and
            Centro de Investigación en Astronomía, Universidad Bernardo O’Higgins, Santiago, General Gana 1760, 8370993, Chile\\
             }

\titlerunning{Density and velocity correlation in ISM}
\authorrunning{K.H. Yuen, K.W. Ho \& A. Lazarian}
\begin{document}
%\label{firstpage}
%\pagerange{\pageref{firstpage}--\pageref{lastpage}}

% These dates will be filled out by the publisher
\date{Accepted XXX. Received YYY; in original form ZZZ}

% Enter the current year, for the copyright statements etc.
%\pubyear{2022}

% Don't change 

% Abstract of the paper
\abstract
{The existence of magnetized turbulence in the interstellar neutral hydrogen (HI) is well accepted. On the basis of theory, a number of techniques to obtain turbulence spectrum and magnetic field direction and strength have been developed and successfully applied to HI spectroscopic data. To better separate the imprints of density and velocity fluctuations to the channel maps, a new theory-based technique, the Velocity Decomposition Algorithm (VDA, \citealt{VDA}), has been created. The technique demonstrates that the intensity fluctuations are separated into a component $p_v$ that {\it mostly} arise from velocity fluctuations and $p_d$ that {\it mostly} arise from density fluctuations. The VDA helps to clarify the nature of the filamentary structure observed in channel maps.}
{A recent publication \citep{K22} claims that the application of VDA to HI4PI data provides a negative correlation of $p_v$ and $p_d$, which according to the authors invalidates the technique since it requires that of $p_v$ and $p_d$ have zero correlation. However, the quantities $p_v$ and $p_d$ given by VDA are naturally orthogonal which can be trivially checked analytically or numerically. That means the {\it correct} application of the VDA to {\it any data} must provide zero correlation. This is the point that we clarify in this paper and search for the cause of the mistake in the application of the VDA in \citep{K22} that resulted in the erroneous conclusion.  }
{We prove analytically that {\it by construction} $p_v$ and $p_d$ are not correlated.  We identify the likely mistake in the VDA expression that \cite{K22} used and reproduce their figures with the {\bf incorrect} expression.}
{We find that 14 out of 15 figures in \cite{K22} are invalid, and thus their criticism of the VDA is ill founded and arises from their use of incorrect expressions. }
{\bf We conclude that the detrimental mistake that \cite{K22} made at their analysis completely invalidate their scientific claim that \cite{VDA} is not compatible to observation.}

% Select between one and six entries from the list of approved keywords.
% Don't make up new ones.

 \keywords{ISM: clouds – ISM: structure – dust, extinction – turbulence – ISM: magnetic fields – magnetohydrodynamics (MHD)}
\maketitle

%%%%%%%%%%%%%%%%%%%%%%%%%%%%%%%%%%%%%%%%%%%%%%%%%%

%%%%%%%%%%%%%%%%% BODY OF PAPER %%%%%%%%%%%%%%%%%%

\section{Introduction}

% Introducing HI
The atomic neutral hydrogen (HI) is well known to be filamentary and to extend over multiple scales. HI observations became the peril of astrophysical research in the $21^{th}$ century. The physical nature of HI is needed to understand why HI is filamentary and preferentially parallel to magnetic field as revealed by recent observations \citep{Clark15,YL17a} and simulations.  Emission line observations like HP4PI \citep{kh15}, GALFA \citep{P18}, THOR-HI \citep{Beuther16,W20}, FAST (e.g. \cite{Li21}) have advanced our understanding of HI physics including its relation to magnetic field \citep{Clark15,YL17a} and its connection to underlying molecular phase \citep{Kritsuk17}.

A proposed physical understanding of the nature of HI is that some, not all nor none, of the features exhibited in spectroscopic position-position-velocity (PPV) channel maps are formed due to the imprint of turbulence motions along the line of sight. This effect, which is called velocity caustics \citep{LP00}, has been studied theoretically and numerically (\citealt{VDA}). Originally caustics were believed to be highly theorized objects that were not detectable in channel maps yet contributed to their observed intensity. In \citeauthor{VDA} (\citeyear{VDA}) we provided a way for separating the fluctuations of intensity $p=p({\bf X},v)$ at some POS position ${\bf X}$ and velocity $v$ into velocity fluctuations $p_v({\bf X},v)$ and density fluctuations $p_d({\bf X},v)$. {\it By construction}, $p({\bf X},v)=p_v+p_d$ and $\langle p_v p_d \rangle=0$\footnote{Where, for the sake of simplifying the notations, we do not show the dependence of $p_v$ and $p_d$ on PPV coordinates (${\bf X},v)$}. Our numerical study in \cite{VDA} provided a framework to systemically show that $p_v \neq 0$ in simulations and observational data, and to show that $p_v$ can be more present than $p_d$ in channel maps, contrary to the widely circulated claim that velocity caustics are not measurable in HI data \citep{Clark19,kk20}. 

The goal of this paper is to rebut the central claim of \cite{K22}. First of all,  {\bf by construction of \cite{VDA}}, $\langle p_v p_d \rangle=0$, regardless of any actual physics and whether velocity or density are actually correlated in PPP space. Any opposite results simply mean that the VDA equations are used incorrectly. In what follows, in \S 2 we identify a mistake in VDA expressions that  \cite{K22} use and provide a new analysis of the same data with the correct VDA expressions. Naturally, we get  $\langle p_v p_d \rangle=0$. In \S 3 we offer some background to the history of the controversy and our views on velocity caustics. Our conclusions are presented in \S 4. 

\section{Mistake in VDA equation in Kalberla et al. (2022)}
\label{sec:k22}
\subsection{Algebraic proof of $\langle p_v p_d \rangle=0$}

The main idea in \cite{VDA} is that, for a given spectroscopic cube $p({\bf X},v)$ and its integrated intensity map $I = \int dv p$, the density fluctuations $p_d$ and velocity fluctuations $p_v$ can be given by:
\begin{equation}
\begin{aligned}
  p_v &= p - \left( \langle pI\rangle-\langle p\rangle\langle I \rangle\right)\frac{I-\langle I\rangle}{\sigma_I^2}\\
  p_d &= \left( \langle pI\rangle-\langle p\rangle\langle I \rangle\right)\frac{I-\langle I\rangle}{\sigma_I^2}
\end{aligned}
\label{eq:ld2}
\end{equation}
where $\sigma_I^2 = (I-\langle I\rangle)^2$. Readers can refer to \S 3 of \cite{VDA} for the construction and justification of the technique.
% We shall discuss a specific case in later section (\S \ref{sec:chris}).

% show that mathematically the VDA will always enforce $\langle p_d p_v \rangle = 0$ 

Notice that the use of Eq.(\ref{eq:ld2}) will mathematically guarantee that $\langle p_d p_v \rangle = 0$:
\begin{equation}
    \begin{aligned}
        &\langle p_d p_v \rangle \\
        &= \Big\langle \left(\left\langle pI\rangle-\langle p\rangle\langle I \rangle\right)\frac{I-\langle I\rangle}{\sigma_I^2}\right) \left(p - \left( \langle pI\rangle-\langle p\rangle\langle I \rangle\right)\frac{I-\langle I\rangle}{\sigma_I^2}\right)\Big\rangle \\
        &= \left(\langle pI\rangle-\langle p\rangle\langle I \rangle\right)\frac{\langle pI-p\langle I\rangle\rangle}{\sigma_I^2}-\Big\langle \left(\left\langle pI\rangle-\langle p\rangle\langle I \rangle\right)\frac{I-\langle I\rangle}{\sigma_I^2}\right)^2\Big\rangle\\
        &= \frac{\left(\langle pI\rangle-\langle p\rangle\langle I \rangle\right)^2}{\sigma_I^2}-\frac{\left(\langle pI\rangle-\langle p\rangle\langle I \rangle\right)^2}{\sigma_I^2} = 0
    \end{aligned}
    \label{eq:proof}
\end{equation}
where we utilize the fact that $\sigma_I^2 = (I-\langle I\rangle)^2$ at the 2nd term of line 3. We can see from Eq.\ref{eq:proof} that $\langle p_d p_v \rangle=0$ follows from equation (Eq.\ref{eq:ld2}) {\it Thus if one uses the correct form of Eq.\ref{eq:proof} the orthogonality of $\langle p_d p_v \rangle$ must be present.} 

That also means, {\bf if {\it any data} with or without density and velocity correlations does not reproduce this result, this must be a mistake in the use of the original VDA equations employed, independent of the type of physical models that HI has on the sky.} Notice that all the criticism of the VDA in \cite{K22} based on the fact that in their analysis $\langle p_d p_v \rangle \ne 0$ is invalid. Therefore it is apparent that $\langle p_d p_v \rangle \neq 0$ cannot be obtained with {\bf any} data.

\subsection{Tracking the source of the mistake via reverse engineering}

It is an interesting question how \cite{K22} got their results while using the VDA. Below we attempt to identify the source of their mistake.

We suspect that the error in \cite{K22} is due to the use of incorrect equations:\footnote{Our guess was based on the fact that Fig.1 of \cite{K22} has weird color bar ranges. Usually for $p_v$ the color bar is symmetric, i.e. it should range from $-A$ to $A$ for some value of $A$ as $\langle p_v \rangle=0$. However, we spot that the value of $p_v$ is abnormally high and that of $p_d$ is abnormally low in \cite{K22}, suggesting that there are over-subtraction in the term $\langle pI\rangle -\langle p\rangle\langle I\rangle$. We tried a few combinations, e.g. $\langle pp\rangle -\langle p\rangle\langle I \rangle$ and $\langle pI\rangle - \langle I\rangle\langle I\rangle$ and finally realize that Eq.\ref{eq:ld_kalberla} is the best in reproducing Fig.1 of \cite{K22}. See \url{https://www.github.com/kyuen2/kalberla_2022} for the illustration. }
\begin{equation}
\begin{aligned}
  p_{v,kalberla22} &= p - \left( \langle {\color{red} p} \rangle-\langle p\rangle\langle I \rangle\right)\frac{I-\langle I\rangle}{\sigma_I^2}\\
  p_{d,kalberla22} &=\left( \langle {\color{red} p} \rangle-\langle p\rangle\langle I \rangle\right)\frac{I-\langle I\rangle}{\sigma_I^2}.
\end{aligned}
\label{eq:ld_kalberla}
\end{equation}

To test whether our identification of mistakes in \cite{K22} is right we apply Eq.\ref{eq:ld_kalberla} to the same regions used in \cite{K22}, namely (1) full sky HI4PI and the three examples \cite{VDA} have used: (2) the $8^o \times 8^o$ GALFA data centered at RA=4.00 and DEC=10.35; (3) the $8^o \times 8^o$ GALFA data centered at $RA=228.00^o$ and $DEC=18.35^o$; (4) and the region considered in \cite{Clark15}, i.e. $195^o<RA<265^o, 19.1^o<DEC<38.7^o$.

\subsection{Reproduction of Figures in Kalberla et al. (2022) with erroneous formulae}

\begin{figure*}
    \centering
    \includegraphics[width=0.90\textwidth]{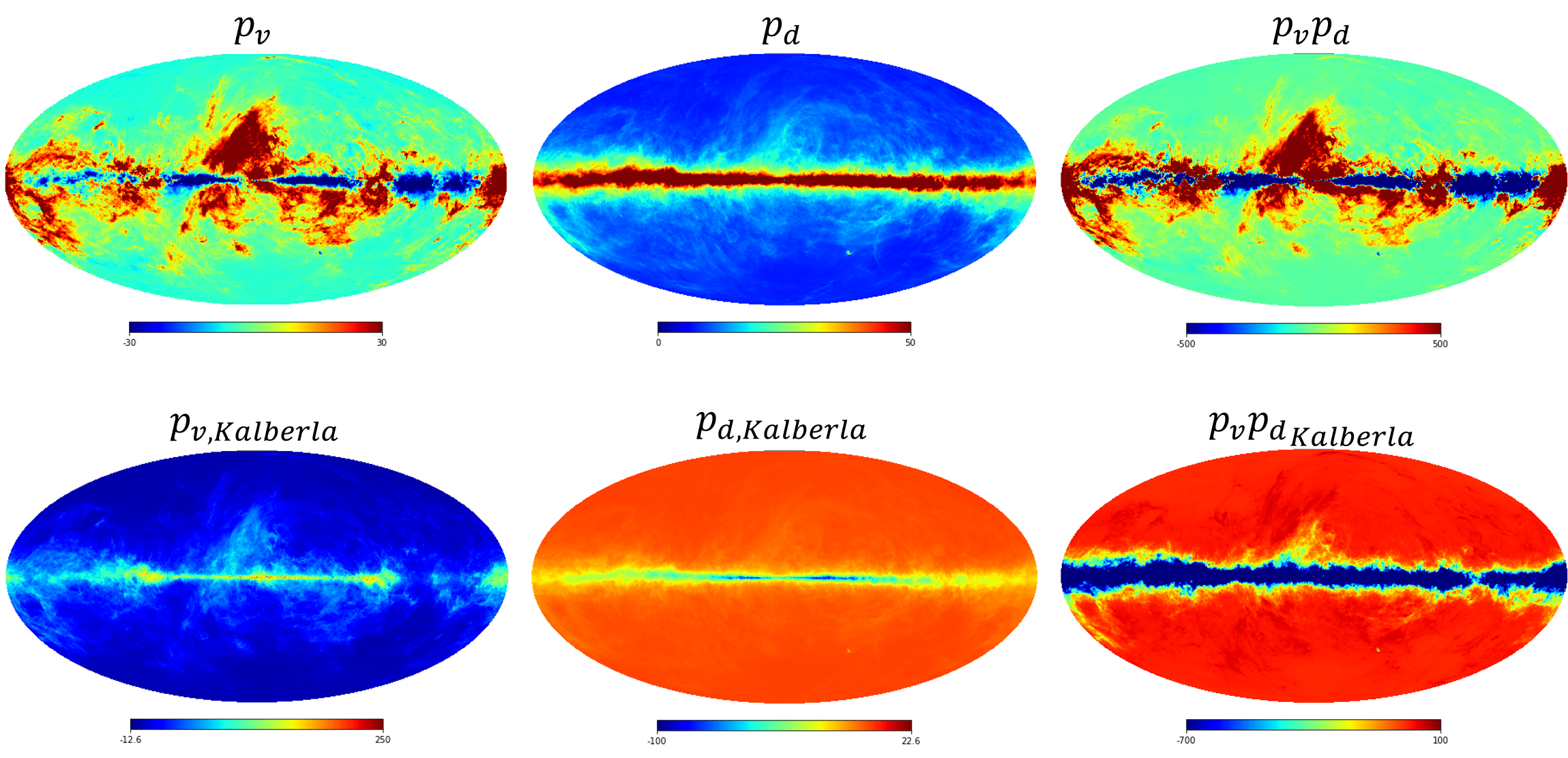}
    \caption{A set of figures showing the $p_d,p_v,\langle p_dp_v\rangle$ through Eq.\ref{eq:ld2} (upper row) and Eq.\ref{eq:ld_kalberla} (lower row). Readers should notice that the lower row of the paper is visually very similar to that of \cite{K22}'s Figure 1, as we intentionally use very similar colorbar as theirs.}
    \label{fig:imshow}
\end{figure*}

To start with, we use the full sky HI4PI data cube. The public data from HI4PI website that we use has a channel width of $\Delta v = 1.29 km/s$, so we are constrained to select the channel $v_{lsr} \approx 1.39km/s \pm 0.65$ to reproduce their results\footnote{We believe that the channel resolution ($\Delta v=1km/s$) quoted in \S 4 of \cite{K22} is an apparent typo. See Table 1 of HI4PI collaboration for the correct number}.  { In Fig. \ref{fig:imshow} we show the actual $p_d, p_v$ using the correct VDA formula given by Eq. (1) and given by Eq. (2). When we use the incorrect formula given by Eq. (\ref{eq:ld_kalberla}) our results are very similar to \cite{K22}'s Figure 1. The results with Eq. (1) differs significantly from \cite{K22}'s Figure 1.}

\begin{figure*}
    \centering
    \includegraphics[width=0.45\textwidth]{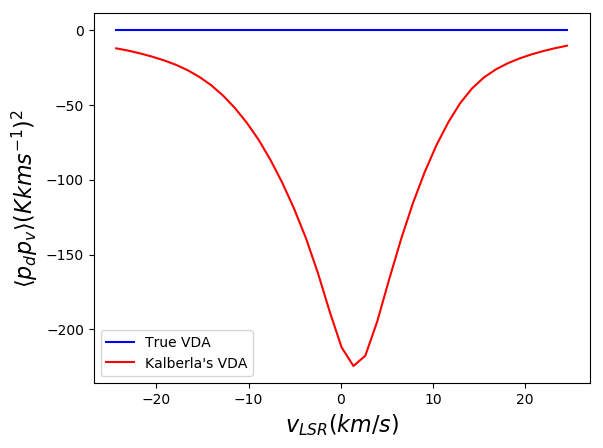}
    \includegraphics[width=0.45\textwidth]{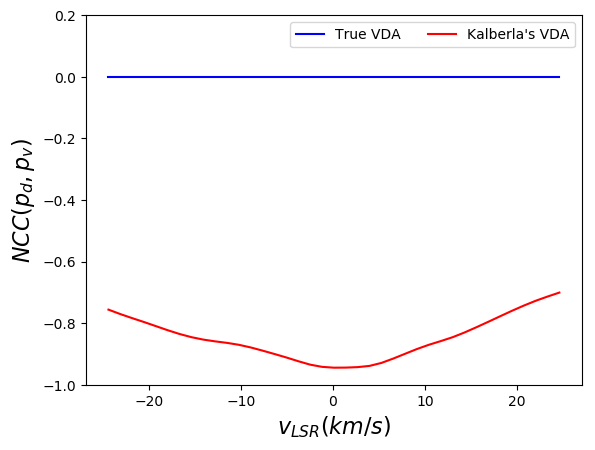}
    \caption{Two figures showing how the $\langle p_dp_v\rangle$ (in units of $(K km s^{-1})^2$ and $NCC(p_d,p_v)$ varies as a function of $v_{lsr}$ (in $km/s$) for the true VDA equation and the suspicious equation used in  \cite{K22} . We can see from the two figures that our VDA method (Eq.\ref{eq:ld2}) correctly reproduce the expected orthogonality according to Eq.\ref{eq:proof}. Nevertheless, we obtain something very similar to their results (highly negative $\langle p_dp_v\rangle$). }
    \label{fig:kalberla_fig23}
\end{figure*}

We then focus on Fig.\ref{fig:kalberla_fig23} of \cite{K22}. The left and middle panels of Fig.\ref{fig:kalberla_fig23} shows both $\langle p_d p_v\rangle$ and $NCC(p_d,p_v)$ using both sets of equations (Eq.\ref{eq:ld2}, Eq.\ref{eq:ld_kalberla}). We can see that the actual correct formula (Eq.\ref{eq:ld2}) correctly reproduce the expected $0$ for both $\langle p_d p_v\rangle$ and $NCC(p_d,p_v)$. 

However, if we use Eq. (\ref{eq:ld_kalberla}) employed in \cite{K22}, we retrieve the pattern that is shown in Fig.2 of \cite{K22} as in the left of Fig.\ref{fig:kalberla_fig23}. We can see that due to the mistake, \cite{K22} cannot reproduce the default orthogonality condition for $p_v$
and $p_d$. As a result, the corresponding $NCC(p_d,p_v)$ is also non-zero as we show in Fig.\ref{fig:kalberla_fig23}\footnote{Notice that the amplitude of $\langle p_d p_v \rangle$ that we obtain using Eq.\ref{eq:ld_kalberla} is off by a factor of 2 compared to Fig. 2 of \cite{K22}. However, we believe this is due to the fact that we do not have access to their HI data with channel width $\Delta v=1km/s$, as the publicly available cube has $\Delta v=1.29km/s$. The reasoning is very simple: A lower value of $p$ will significantly decrease the value of $p_{d,kalberla}$. Naturally, the thickness of the channels cannot affect the default condition $\langle p_d p_v \rangle=0$ }.

In addition, we compute the NCC curve using both the correct and erroneous formula on data sets (2)-(4) as discussed above. We see in Fig 3 that our VDA formula clearly reproduce $\langle p_d p_v \rangle=0$ as expected by Eq.\ref{eq:proof}. Again, we found that using the erroneous formula (Eq.\ref{eq:ld_kalberla}) we could reproduce the results of Fig 4 upper, middle, and lower panel of \cite{K22}. 

All the figures beyond Fig.4 of \cite{K22} are based on the same erroneous equation used in Fig 1,2 and 4, and are all therefore \footnote{Fig.3 of \cite{K22} is the only figure that is irrelevant to VDA as it was meant to explain Fig 2 of \cite{K22}. However, Fig.2 of \cite{K22} is wrong.} invalid.

\begin{figure*}
    \centering
    \includegraphics[width=0.33\textwidth]{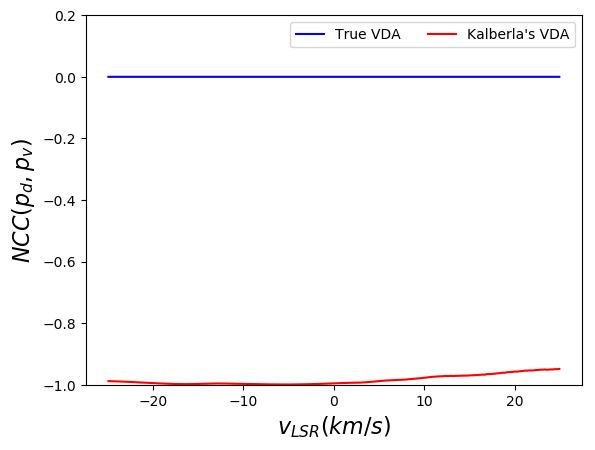}
    \includegraphics[width=0.33\textwidth]{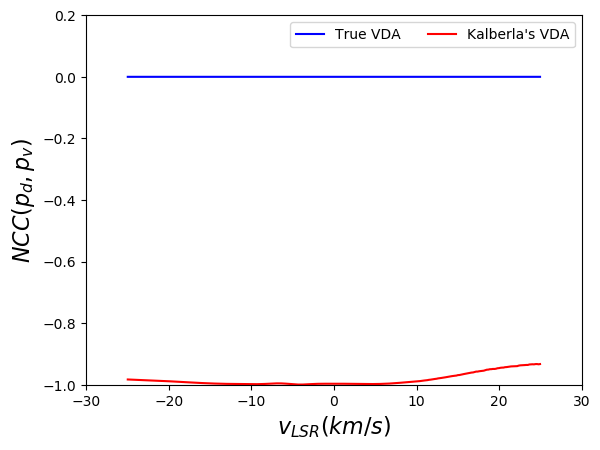}
    \includegraphics[width=0.33\textwidth]{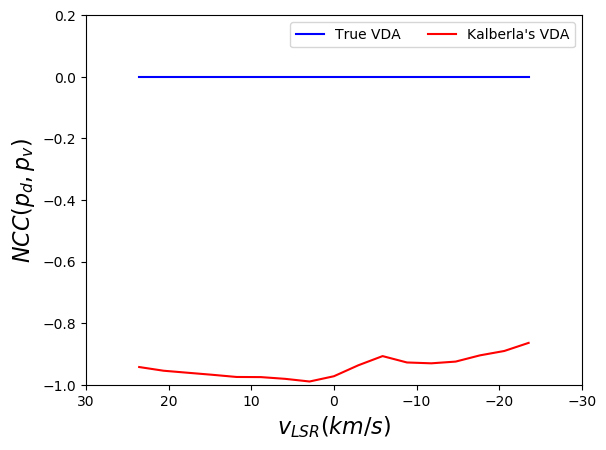}
    \caption{Three figures showing the NCC values as a function of $v_{lsr}$ (in $km/s$) from the actual Eq.\ref{eq:ld2} and the wrongly used Eq.\ref{eq:ld_kalberla} for the three regions: left: $8^o \times 8^o$ GALFA data centered at RA=4.00 and DEC=10.35; middle: to the $8^o \times 8^o$ GALFA data centered at $RA=228.00^o$ and $DEC=18.35^o$; right: for the region $195^o<RA<265^o, 19.1^o<DEC<38.7^o$. } 
    \label{fig:GALFA_cube}
\end{figure*}

\section{Discussions}
\label{sec:discussion}

{\bf Our tests and results presented in the previous section are significant enough that we could end the present publication with the following conclusion}.  It is rather apparent that \cite{K22} made a rudimentary mistake when implementing the VDA method, which is the foundation of their entire paper.  Our current letter is intended to clarify that the harsh criticism of the VDA in \cite{K22} is based on the erroneous use of the VDA equation. Therefore we focus on this central result of \cite{K22} and do not comment on other issues, e.g. whether Gaussian decomposition is applicable on top of VDA. We note that \cite{K22} is similar to \cite{kk20} which provided a rather harsh criticism of the LP00 theory and its applications. In \cite{VDA}, we answered of the criticism in \cite{kk20}. Here we find that it is necessary and important for the community to understand some of the history and our views concerning velocity caustics.

While dismissing the VDA, \cite{K22} strongly supported the interpretation that HI filaments are {\it entirely} density fluctuations  (see \cite{Clark19,kk20}). The controversy over the importance of velocity caustics in HI maps started with \cite{Clark19} argument that there is a strong correlation between the FIR dust emission and the intensity fluctuations of HI channel maps. This argument was supported in a number of publications, e.g. \citealt{2019ApJ...886L..13P,kk20}). The arguments that it is wrong to disregard the velocity caustics were also presented. For example \cite{reply19} points out that the results from \cite{Clark19} were not properly normalized and the correlation of FIR dust emission and HI channel maps does not imply that there is no correlation between velocity fluctuations and FIR dust emissions\footnote{This is a trivial logic. Showing A is similar to B does not exclude the possibility that C can also be similar to B despite A and C being mutually exclusive. Here A is HI channels, B is FIR dust emission and C is caustics. Similarity operator is not associative.}. \cite{VDA} provided further evidence of the role of velocity caustics, proving, at the very least, that the claim that {\it all structures on the sky are predominately density fluctuations is wrong}. In fact, the relative contribution of density and velocity depends on the region of study. For example, in \cite{VDA} we gave two examples, namely the regions (2, RA=4, DEC=10.35) and (3, RA=228, DEC=18.35) that we captured from GALFA-DR2. The region (2) under our analysis has subdominant density contribution (Fig.22 of \citealt{VDA}). However, for region (3) we actually found from our analysis that the density and velocity contributions are in equal portion for that particular region (Fig 23 of \citealt{VDA}). We believe that in accordance with LP00 theory, both density and velocity fluctuations contribute to PPV fluctuations and their relative role could be evaluated by the toolbox in \cite{VDA}. \cite{VDA} demonstrates that velocity caustics provide significant contribution to observed filaments in PPV, which is contrary to the claim from \cite{Clark19,kk20}. 

It is also important to note that the concept of velocity and density contributions in the channel maps does not necessarily extend to the full sky. For instance, the application of the VDA to a big statistically inhomogeneous region can induce uncertainties. It is a mistake, therefore, to use the VDA on the full sky since clearly the galactic center is not one self-consistent turbulent system as are isolated molecular clouds\footnote{Which we found that the authors of \cite{K22} actually knew this fact (p14 second last paragraph of their paper) despite they still decide to apply VDA to the full sky.}. Only when the MHD approximation start to take place, i.e. scales smaller than the injection scale (\citealt{CL09,spectrum}) should the VDA technique be applicable\footnote{the injection scale in the ISM is commonly thought to be \textasciitilde{}100 pc \cite{CL09} though this value is still in debate in the community}. However, the applicability of VDA on scales beyond injection scale by itself is a topic of study. {\bf \cite{K22} not only mistake the central equation of the VDA but use the technique in a region where it does not apply, questioning whether they understand its basic theory.}

%{\bf Readers should not make mistakes like \cite{K22} that attempted to apply some technique (wrongly) to some unreasonable region and claim that someone's technique is incorrect. }

\cite{K22} also questioned the validity of the MHD simulations. For instance, they claim "{\it Our result are in clear contradiction to the MHD simulations and theoretical expectations from uncorrelated velocity and density fields with $NCC \sim 0$.}" (Sec 6.5 of \citealt{K22}). Given their critical mistakes in their paper, we abstain from replying to their criticisms, which should be merely regarded as a biased opinion with no solid ground. 

%As we discussed above, VDA by itself is a self-consistent framework. 
\section{Summary}
\label{sec:conclusion}
In this paper, we show that \cite{K22} made a simple yet fatal mistake in terms of the implementation of Eq.\ref{eq:ld2} when analyzing the observational data. This mistake is not a scientific one, but rather a rudimentary programmatic error leading to the erroneous main results of \cite{K22} and invalidating all their figures but Figure 3. {\bf Their discussions based on their wrongly computed $p_v$ leading to the erroneous "discovery" that $\langle p_d p_v \rangle \neq 0$} is also misleading. In particular, 
\begin{itemize}
    \item We showed that $\langle p_d p_v \rangle = 0$ is always true if $p_v$ and $p_d$ are defined by the VDA equations. Therefore \cite{K22} finding that $\langle p_d p_v \rangle \neq 0$ is a result of incorrect use of the VDA equations.
    \item We identified the mistake in equations employed in \cite{K22} and reproduced their figures analyzing their data using erroneous equations. In parallel we provided the results obtained with the correct formulae and demonstrated that we can indeed obtain $\langle p_dp_v\rangle=0$ in any selected HI4PI data.
    \item Furthermore, the authors of \cite{K22} mistakenly use VDA on the full sky HI data. This reveals a lack of understanding of the core the VDA technique. The VDA is designed to work for a single turbulent system. It is not meant to apply to full-sky HI data with a huge statistical inhomogeneity that arises from, e.g. galactic motions\footnote{Notice that the authors of \cite{K22} suggest that there are multiple clouds along the line of sight. The correct way of doing VDA analysis here is to first identify the clouds' individual Gaussian profile, apply VDA on those individual profile, and cross check with our analysis.}.
\end{itemize}

\begin{acknowledgements}
KHY and KWH thank Chi Yan Law in timely reminding us the arXiv paper from \cite{K22}, and provide valuable comments on our paper. KHY and KWH also thank Avi Chen in providing valuable comments and suggestions to our manuscript. KHY KWH, \& AL acknowledge the support the NSF grant AST 1212096 and NASA grant NNX14AJ53G.  All the data used are publicly available (HI4PI: \cite{kh15}, GALFA-DR2: \cite{P18}). The exact data that the author processed, including the intermediate product and the end results, will be shared on reasonable request to the corresponding author.The Velocity Decomposition code in {\it Julia} (\citealt{VDA}) is publicly available and  can be found at \url{https://www.github.com/kyuen2/LazDDA}. The workflow of the current paper can be found at \url{https://www.github.com/kyuen2/kalberla_2022/}.
\end{acknowledgements}

\end{document}